\documentclass[apm,reprint,amsmath,amssymb]{revtex4-1}

\usepackage{graphicx}
\usepackage{bm}

\begin{document}

\title{Tuning Optical Properties of
Transparent Conducting Barium Stannate by Dimensional Reduction}

\author{Yuwei Li$^{1,2}$}
\author{Lijun Zhang$^3$}
\author{Yanming Ma$^1$}
\author{David J. Singh$^4$}

\affiliation{$^1$State Key Lab of Superhard Materials, Jilin University, 130012,
Changchun, China \\
$^2$Beijing Computational Science Research Center, Beijing 100084, China \\
$^3$College of Materials Science and Engineering, 
Jilin University, 130012, Changchun, China \\
$^4$Materials Science and Technology Division,
Oak Ridge National Laboratory, Oak Ridge, Tennessee 37831-6056}

\date{\today}

\begin{abstract}
We report calculations of the electronic structure and optical
properties of doped $n$-type perovskite BaSnO$_3$ and layered
perovskites.
While doped BaSnO$_3$ retains its transparency
for energies below the valence to conduction band onset,
the doped layered compounds exhibit below band edge optical
conductivity due to transitions from the lowest conduction band.
This gives absorption in the visible
for Ba$_2$SnO$_4$. Thus it is important to minimize
this phase in transparent conducting oxide (TCO) films.
Ba$_3$Sn$_2$O$_7$ and Ba$_4$Sn$_3$O$_{10}$
have strong transitions
only in the red and infrared, respectively.
Thus there may be opportunities for using these
as wavelength filtering TCO.
\end{abstract}


\maketitle


Transparent conducting oxides (TCO) are widely used for electrodes
in solar energy and display technologies. \cite{ginley,kawazoe}
The most commonly used material is
Sn doped In$_2$O$_3$, known as ITO.
This is a stable material that can be
readily deposited in thin films and has excellent TCO performance.
However, there is interest in alternatives to ITO both because
of a desire to reduce the use of In and to identify
TCO materials with a range of other properties and functionalities.
\cite{minami}
For displays, high DC conductivity, low haze and high transparency for the
operating wavelength and viewing angle ranges are needed.
Transparency outside these ranges is not needed and
is often undesirable since
heating due to infrared transmission and ultraviolet photons can
degrade devices, e.g. in outdoor use.
We note that transparency below the absorber band gap is undesirable
in solar photovoltaic applications.
While light filters can be incorporated in various layers of a device
architecture, transparent conductors that also act as filters may be useful.

Perovskite
structure BaSnO$_3$ doped by trivalent cation substitutions for
Ba (e.g. La) and by Sb alloying on the Sn site
shows promise as an ITO alternative.
\cite{cava,singh-basno3,upadhyay,hadjarab,wang-basno3,kim,luo,liu2,kim2,
mizoguchi,mizoguchi-ba,fan,liu,singh3,mun,wadekar,sallis,scanlon,kim3}
Importantly, recent studies of high mobility La doped BaSnO$_3$ transparent
conducting films show transport properties that are still dominated by
extrinsic defects, specifically dislocations.
\cite{kim2,mun,wadekar}
The implication is that there remains considerable room
for even better conductivity with improvements in film perfection.
Recent studies suggested that $n$-type BaSnO$_3$ may have substantial
tunability, e.g. in the band gap via alloying and strain,
\cite{mizoguchi-ba,fan,liu,singh3}
but applications of strain tuning of BaSnO$_3$
have yet to be demonstrated.
Here we show that it is possible to tailor the spectral transmission
of BaSnO$_3$ by a dimensional reduction strategy.


We report first principles calculations for
doped BaSnO$_3$, in relation to 
members of the Ruddlesden-Popper homologous series
derived from BaSnO$_3$ by periodic insertion of nominally
insulating BaO layers.
The first two members of the series
are known bulk equilibrium stable phases
in the BaO -- SnO$_2$ pseudobinary phase diagram
that are readily made using common methods.
\cite{wagner,hinatsu,green,kennedy,ropp,yamashita,stanulis}
Perhaps other phases can also be grown, in analogy
with the Sr-Ti-O system. \cite{lee}
This is also consistent with the Sr-Sn-O system, which has a known
Ruddlesden-Popper series. \cite{green2,fu}
Kamimura and co-workers have recently shown that members of the Sr-Sn-O
Ruddlesden-Popper series can show very different luminescence properties
when activated. \cite{kamimura}
Furthermore, the members of the Ba-Sn-O
Ruddlesden-Popper series have good in plane
lattice matches with each other and are presumably compatible with
each other in thin film form.

The experimental lattice parameters and spacegroups are
BaSnO$_3$, cubic, $Pm3m$  $a$=4.116 \AA,
\cite{vegas,maekawa,bevillon}
Ba$_2$SnO$_4$, body centered tetragonal, $I4/mmm$,
$a$=4.1411 \AA, $c$=13.2834 \AA, \cite{green} and
Ba$_3$Sn$_2$O$_7$, body centered tetragonal, $I4/mmm$,
$a$=4.129 \AA, $c$=21.460 \AA.
\cite{hinatsu}
As discussed below, we also did calculations for the next member
of the series, Ba$_4$Sn$_3$O$_{10}$, which is a phase that has
not been reported experimentally. For this we did calculations
both using a fully relaxed structure and a partially relaxed structure
in which the atomic coordinates were relaxed but the lattice parameters
were fixed to interpolated experimental values from the known phases.

The main first principles calculations were performed using the general
potential linearized augmented planewave (LAPW) method
\cite{singh-book} as implemented in the WIEN2k code.
We also did calculations to cross-check
using the VASP code with hybrid HSE06 functional
\cite{heyd,heyd1}
and the generalized gradient approximation of Perdew, Burke and Ernzerhof
(PBE-GGA). \cite{pbe}

For the LAPW calculations,
we used well converged basis sets including local orbitals for
the semicore states of the alkaline earth elements, the O 2$p$
state and the Sn $4d$ state. We used the standard LAPW basis sets
rather than the APW+lo method. \cite{sjo}
The LAPW sphere radii were 2.4 bohr, 2.25 bohr and 1.6 bohr for
Ba, Sn and O, respectively, and a basis set cut-off, $k_{max}$,
set by the criterion $R_{min}k_{max}$=7.0 was used, where $R_{min}$
is the O LAPW sphere radius.

We fixed the lattice parameters to the experimental values and relaxed the
internal atomic coordinates of the layered compounds using the
PBE-GGA.
Following the structure optimization, we did
electronic structure and optical calculations using the 
modified Becke-Johnson type potential functional of Tran and Blaha, \cite{mbj}
denoted TB-mBJ in the following.
This potential gives band gaps in remarkably good accord with experiment
for a wide variety of simple semiconductors and insulators,
\cite{mbj,koller,singh1,kim-mbj,singh2}
including perovskite stannates.
\cite{fan}
We compared the TB-mBJ results for the undoped compounds with
HSE06 calculations as discussed below.

\begin{figure}
\includegraphics[width=\columnwidth,angle=0]{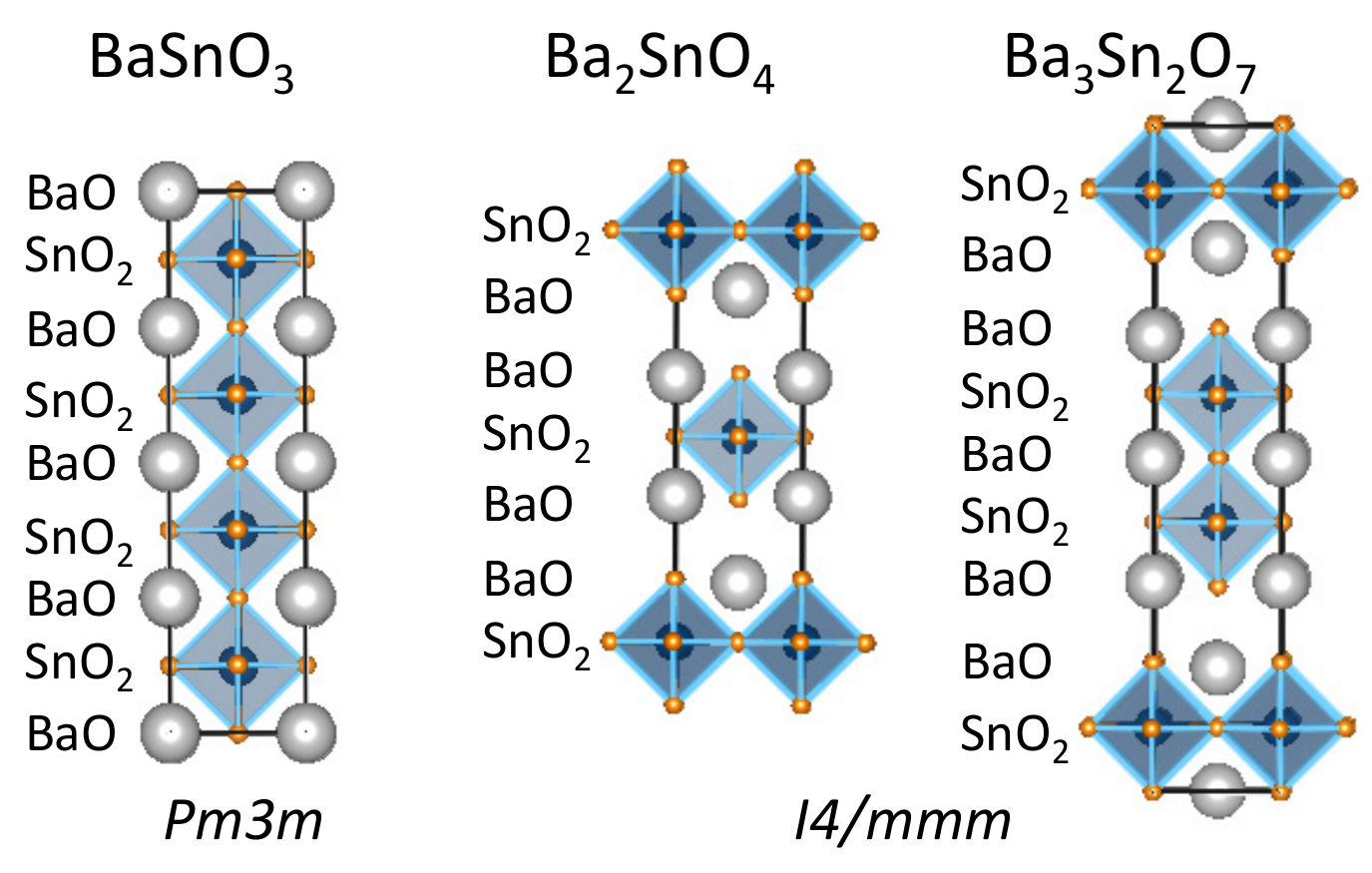}
\caption{Cubic perovskite structure and the first
two members of the Ruddlesden-Popper series, showing the stacking
of BaO and SnO$_2$ layers.}
\label{structs}
\end{figure}

Doping by La was treated using the virtual crystal approximation.
The virtual crystal approximation
is an average potential approximation. 
It goes beyond rigid bands, and specifically
includes composition
dependent distortions of the band structure.
The use of the virtual crystal approximation is supported by the fact
that in these compounds the highly electropositive elements, Ba and La,
are fully ionized and serve only to stabilize the structure and donate
charge to the conducting bands derived from Sn and O.
Ba$_3$Sn$_2$O$_7$ has two crystallographically distinct Ba sites,
{\em i.e.} the $2b$ perovskite-like site between the two SnO$_2$ layers
and the $4e$ rocksalt-like site in between the perovskite blocks (Fig.
\ref{structs}).
For this compound we performed calculations in which we did virtual
crystal with La on site $2b$, on site $4e$ and equally distributed
on the two sites. The results were practically the same, which supports
the validity of the virtual crystal approximation.


There have been a number of prior first principles studies of BaSnO$_3$
as well as the related perovskites, CaSnO$_3$ and SrSnO$_3$.
The Sn $s$ nature of the conduction band and its role in the good
conductivity of $n$-type doped BaSnO$_3$ were established early on.
\cite{singh-basno3}
This basic result was found in subsequent band structure calculations
with various methods.
\cite{kim2,mizoguchi,mizoguchi-ba,fan,liu,singh3,mun,wadekar,sallis,scanlon,kim3}
Large variation in the band gap between BaSnO$_3$, SrSnO$_3$ and CaSnO$_3$,
was found by Mizoguchi and co-workers \cite{mizoguchi}, and subsequently
discussed in terms of strain. \cite{fan,singh3}

\begin{figure}
\includegraphics[width=\columnwidth,angle=0]{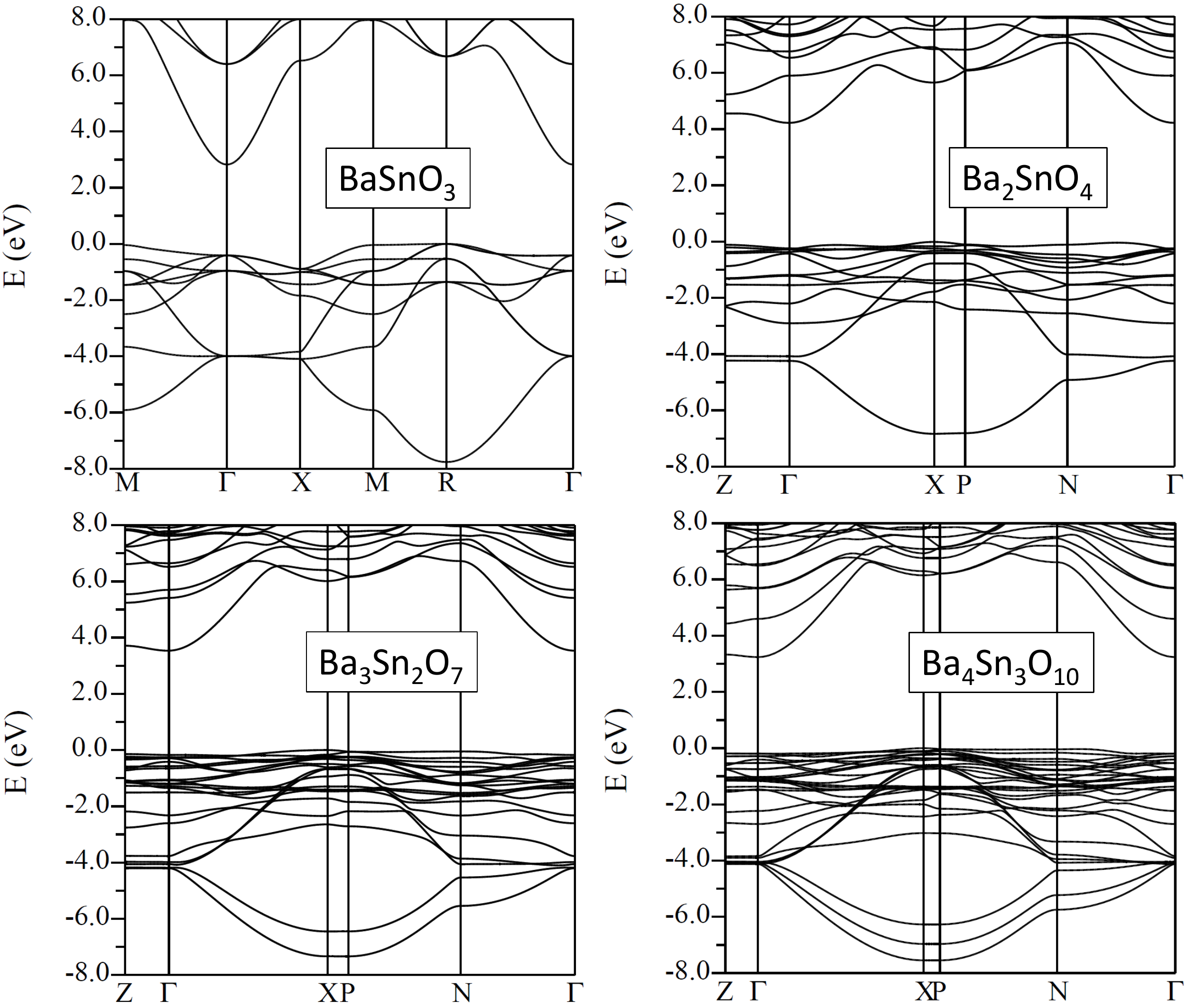}
\caption{Band structures of cubic BaSnO$_3$, and body centered tetragonal
Ba$_2$SnO$_4$, Ba$_3$Sn$_2$O$_7$ and hypothetical Ba$_4$Sn$_3$O$_{10}$.
The band structure of BaSnO$_3$ is that previously reported in
Ref. \onlinecite{fan}, reproduced with permission from
PLoS ONE {\bf 9}, e91423 (2014).
Copyright 2014 Fan, Zheng, Chen and Singh,
an open access work distributed under the terms of the Creative Commons
Attribution License.}
\label{bands}
\end{figure}

The band structures with the TB-mBJ potential are shown in
Fig. \ref{bands}.
The calculated indirect gaps using the TB-mBJ potential are
2.82 eV, 4.22 eV and 3.53 eV for BaSnO$_3$, Ba$_2$SnO$_4$ and
Ba$_3$Sn$_2$O$_7$, respectively.
The calculated TB-mBJ band gap of bulk NaCl
structure BaO, lattice parameter 5.523 \AA, is 3.61 eV, indirect between
$X$ and $\Gamma$.
Two aspects that are immediately apparent from the band structures
are (1) the layered compounds show a dispersive $s$
band at the conduction band minimum similar to BaSnO$_3$ and (2)
the layered compounds show little dispersion in the $c$-axis
direction, so the layering is indeed strongly reflected in the
electronic structure. The band narrowing in the $c$-axis direction
is a kind of quantum confinement effect.

Table \ref{tab-gap} gives the band gaps as obtained with the PBE-GGA
functional, the
TB-mBJ potential and the HSE06 hybrid functional.
As may be seen there is a good agreement between the TB-mBJ
and the HSE06 results, while as usual the PBE-GGA underestimates gaps.

\begin{table}
\caption{Band gaps for the Ba-Sn-O Ruddlesden-Popper compounds
as obtained with the PBE-GGA functional, the
 TB-mBJ potential and the HSE06 hybrid
functional.}
\begin{tabular}{lccc}
\hline
Compound & PBE-GGA (eV) & TB-mBJ (eV) & HSE06 (eV) \\
\hline
BaSnO$_3$ & 0.98 & 2.82 & 2.66 \\
Ba$_4$Sn$_3$O$_{10}$ & 1.61 & 3.24 & 3.21 \\
Ba$_3$Sn$_2$O$_{7}$ & 2.03 & 3.53  & 3.60 \\
Ba$_2$SnO$_{4}$ & 2.83 & 4.22 & 4.41 \\
\hline
\end{tabular}
\label{tab-gap}
\end{table}

\begin{figure}
\includegraphics[width=\columnwidth,angle=0]{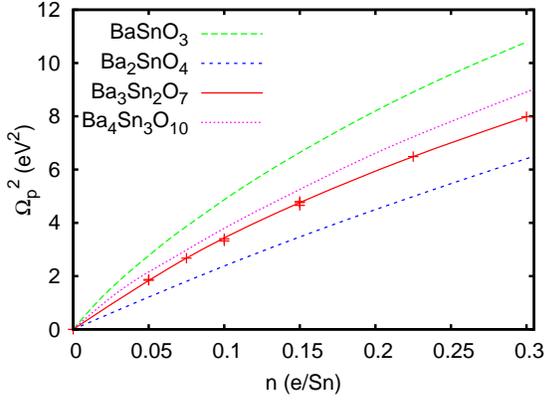}
\caption{In-plane square plasma frequency of
$n$-type BaSnO$_3$,
Ba$_2$SnO$_4$, Ba$_3$Sn$_2$O$_7$
and hypothetical Ba$_4$Sn$_3$O$_{10}$ as a function of carrier concentration
in electrons per Sn atom. Calculations were done
for doping on the different Ba-sites in  Ba$_3$Sn$_2$O$_7$ (see text) with
data indicated by points.}
\label{plasma}
\end{figure}

Fig. \ref{plasma} gives the
in-plane plasma frequencies, $\Omega_p$ for $n$-type doping 
as a function of doping in carriers per Sn.
Conductivity in metals and degenerately doped semiconductors
depends on the plasma frequency, $\sigma \propto \Omega_p^2\tau$
where $\tau$ is an effective inverse scattering rate.
As seen, if one assumes similar scattering rates,
the conductivity follows a perhaps
expected trend following the density of Sn ions ({\em i.e.} falling as the
concentration of insulating BaO increases).
Specifically, for in-plane conduction,
$\sigma$(BaSnO$_3$)$>$$\sigma$(Ba$_3$Sn$_2$O$_7$) $>$$\sigma$(Ba$_2$SnO$_4$).
The volumes per Sn are 69.73 \AA$^3$, 91.47 \AA$^3$ and 113.90 \AA$^3$,
for BaSnO$_3$, Ba$_3$Sn$_2$O$_7$ and Ba$_2$SnO$_4$, respectively.
The BaO layers, while not contributing to conduction, also would
not be expected to contribute to optical absorption, and therefore
the result suggests that the layered compounds could be good
TCO as well.
As discussed below, our optical calculations show
that this may be
the case for Ba$_3$Sn$_2$O$_7$ and higher members of the
series, but is not the case for
Ba$_2$SnO$_4$.

The good TCO behavior of BaSnO$_3$ can be understood
in terms of three main reasons: (1) The material can be effectively
doped $n$-type to high carrier concentrations, (2) It has
a highly dispersive Sn $s$ derived conduction band that leads
to high mobility and (3) there is a large gap
between the lowest conduction band at the
conduction band minimum (CBM), which is at $\Gamma$,
and the next conduction band at $\Gamma$.

\begin{figure}
\includegraphics[width=\columnwidth,angle=0]{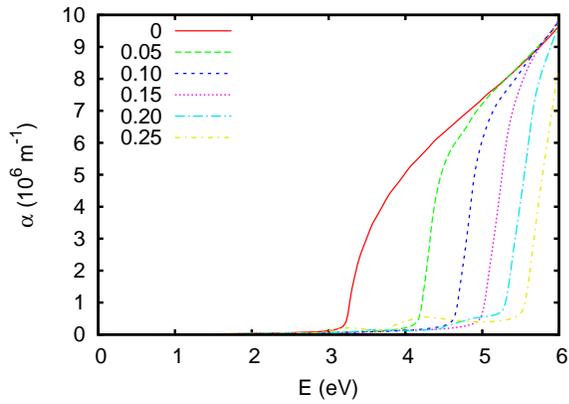}
\caption{Calculated absorption spectra for
$n$-type BaSnO$_3$ as a function of doping level in carriers per Sn.
The data for zero doping are from Ref. \onlinecite{fan}.}
\label{BaSnO3-abs}
\end{figure}

\begin{figure}
\includegraphics[width=\columnwidth,angle=0]{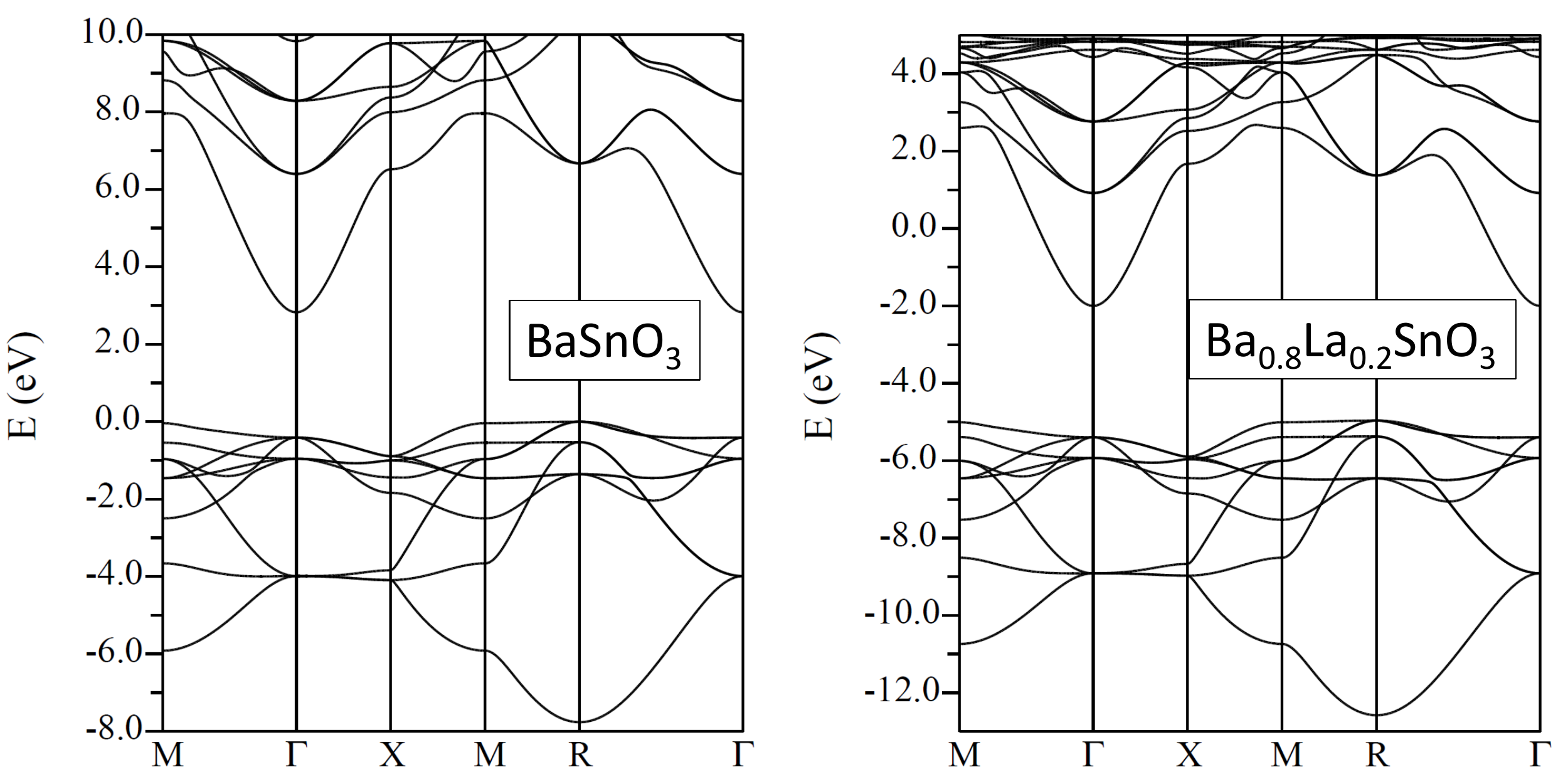}
\caption{Comparison of the band structure of undoped BaSnO$_3$ (left)
and virtual crystal Ba$_{0.8}$La$_{0.2}$SnO$_3$ (right). 
The energy zero is at the valence band maximum for BaSnO$_3$
and at the Fermi level for the doped compound. Note the relatively
minor band distortions.
}
\label{bands-vc}
\end{figure}

The calculated absorption spectra for virtual crystal doped
BaSnO$_3$ are shown in Fig. \ref{BaSnO3-abs}.
As seen, the optical gap increases with carrier concentration. This
is a consequence both of the electrostatic effect of adding electrons
to a band (electrons added to
the Sn $s$ derived conduction band by doping provide
a repulsion to other electrons in this orbital raising the energy of the band),
and the increase in Fermi level within the band as electrons are added.
Here the Fermi level shift is the larger effect. For example, with
0.2 electrons per Sn, we obtain a Fermi level 1.99 eV above the conduction band
minimum in our virtual crystal calculations, while the indirect
gap between the valence band maximum and conduction band minimum increases
by only 0.14 eV. This latter increase is a non-rigid-band effect. In addition
to this band gap there are other small distortions of the band structure.
This is shown in Fig. \ref{bands-vc}.

The increase in apparent optical gap with La doping is possibly
consistent with optical absorption spectra of
BaSnO$_3$ and (Ba,La)SnO$_3$ obtained by Kim and co-workers,
\cite{kim2}
whose experimental spectra show such an increase.
However, in view of the broadening of the reported experimental
spectra with La addition further measurements would be desirable.

The band structures of Ba$_2$SnO$_4$ and Ba$_3$Sn$_2$O$_7$ do not show
large gaps above the CBM, but instead show other bands within 3 eV
of the CBM. These arise from the dimensional reduction, specifically
the strongly reduced hopping along the $k_z$, $c$-axis direction, which narrows
the conduction bands.
Additionally, in Ba$_3$Sn$_2$O$_7$, there are two perovskite layers
in the perovskite block, which leads to a band structure with
pairs of bands in the $k_z$=0 plane that are characterized by
even or odd symmetry
with respect to reflection in the BaO plane between the SnO$_2$ layers.
Such pairs of bands are then strongly connected by dipole transitions for
electric field polarization along the $c$-axis direction, but not for in-plane
polarization, {\em i.e.} not for light propagating along the $c$-axis
direction.

\begin{figure}
\includegraphics[width=\columnwidth,angle=0]{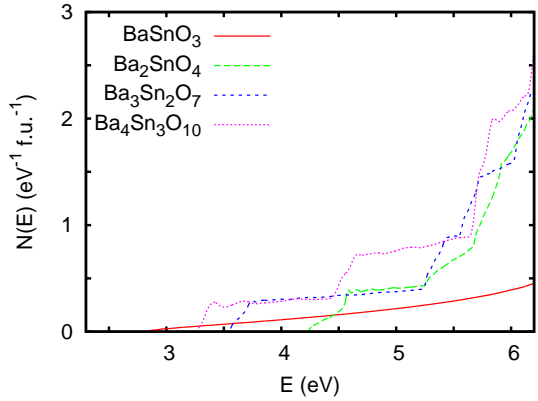}
\caption{Calculated electronic DOS for the conduction
bands. The energy zero is at the valence band maximum. Note the
step like features for all of the layered compounds.}
\label{dos}
\end{figure}

Other consequences of the dimensional reduction are a higher density
of states near the CBM corresponding to the near 2D electronic structure
(note the flatness of the lowest band along the $Z$-$\Gamma$ direction);
a 2D electronic structure has a high, constant density of states near
the band edge, while a 3D band structure has a lower initial density
of states, $N(E)\propto E^{1/2}$, with $E$ relative to the CBM.
This is clearly seen in the densities of states (DOS),
shown for the conduction bands in Fig. \ref{dos}.
In particular, the DOS for BaSnO$_3$ shows a smooth increase with
energy, while prominent steps corresponding to band onsets are present
for the layered compounds.
The consequence of this stepped DOS
is that in 2D the doping dependence of the Fermi level, $E_F$,
will be weaker, with the result that higher doping levels might be
possible, and that following arguments given elsewhere the thermopower
will be strongly enhanced. \cite{hicks,chen} For the conductivity,
$\sigma$ in plane is generically expected to remain similar to the 3D
compound at the expense of a very strong
reduction in $c$-axis conductivity. This means that use of Ba$_2$SnO$_4$
or Ba$_3$Sn$_2$O$_7$ films as conducting oxides would require the production
of textured films.

\begin{figure}
\includegraphics[width=\columnwidth,angle=0]{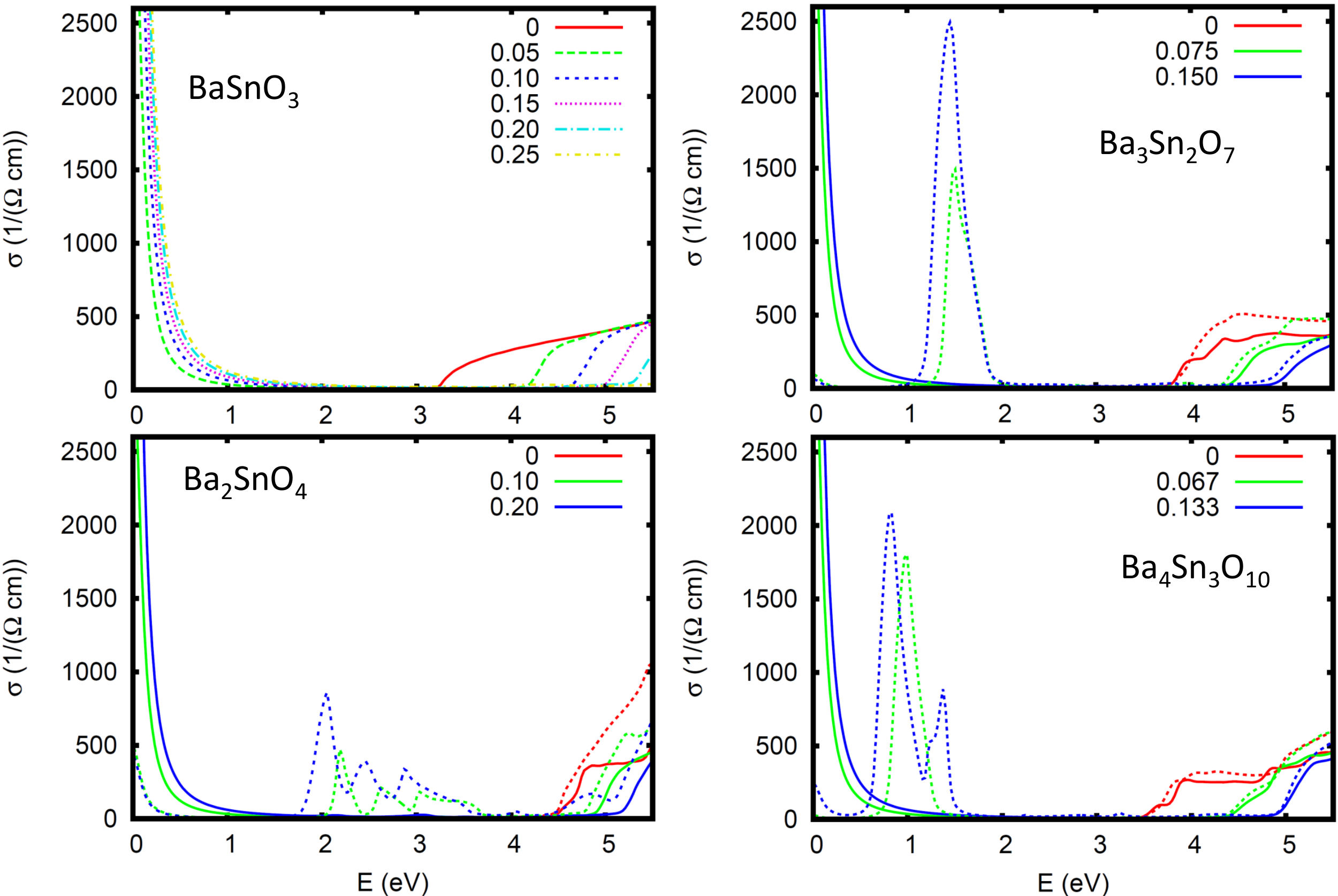}
\caption{Optical conductivities of BaSnO$_3$,
Ba$_2$SnO$_4$, Ba$_3$Sn$_2$O$_7$ and hypothetical
Ba$_4$Sn$_3$O$_{10}$ at various
doping levels in units of electrons per Sn.
For the layered compounds, the solid lines are
for the in-plane, $a$-axis conductivities, while the dashed lines are for
the $c$-axis conductivities.}
\label{cond}
\end{figure}

The optical conductivities of the doped compounds
are given in Fig. \ref{cond}, showing both in-plane and out-of-plane
directions.
These plots include a Drude
contribution, with the Drude weight from the calculated $\Omega_p$
and an assumed broadening, $\gamma$=0.1.
Although undoped Ba$_2$SnO$_4$ is the compound with the highest band
gap, it shows strong optical conductivity (and absorption)
in the visible for light
polarization in the $c$-axis direction, but low optical conductivity
for polarization in the plane, which is the case for normal incidence.
This might suggest that Ba$_2$SnO$_4$ could be used as a transparent
conductor for normal incident light. This may, however, be misleading,
since in reality the disorder would lead to
symmetry breaking, which could then result in some absorption for normal
incident light as well.
In any case, this absorption is a consequence of interband transitions
from the lowest conduction band to higher bands near the zone
center.

Ba$_3$Sn$_2$O$_7$ shows different behavior. It displays very strong
interband transitions for $c$-axis polarized light in the infrared and red
(below $\sim$ 2 eV).
These lead to high peaks in the optical conductivity for $c$-axis
polarization even at moderate doping. This polarization involves
dipole transitions between bands with opposite reflection symmetry
in the plane between the SnO$_2$ layers as mentioned above.
The band making up the CBM
has even reflection symmetry, with an onset of odd
reflection symmetry states 1.7 eV above the CBM.
Importantly,
Ba$_3$Sn$_2$O$_7$ does not show significant absorption in the visible
above 2 eV
for either light polarization, but would be expected to be
strongly absorbing below 2 eV for electric field
out of plane and in the presence of disorder for both polarizations.
More generally, the insertion of BaO layers into BaSnO$_3$ breaks
the translational symmetry along the $c$-axis direction. This then
allows dipole transitions for electric field polarization along the $c$-axis
since the bands folded back to $\Gamma$ are from different original
crystal momenta along $k_z$ (note that the dipole operator is equivalent
to a momentum operator).

As mentioned, we also did calculations for hypothetical Ba$_4$Sn$_3$O$_{10}$,
which is the next member of the homologous series. For this, we used lattice
parameters based on interpolation using the experimental lattice
parameters of BaSnO$_3$, Ba$_4$SnO$_4$ and Ba$_3$Sn$_2$O$_7$.
We then fully relaxed the internal atomic coordinates within this cell.
The lattice parameters 
used were from interpolation, $a$=4.125 \AA, and $c$=29.637 \AA.
Full relaxation with the PBE-GGA yielded slightly larger lattice parameters
of $a$=4.202 \AA, and $c$=30.170 \AA,
also with relaxation of the internal atomic
coordinates. The calculated band gaps are 3.24 eV
for the interpolated structure
and 2.64 eV
for the larger volume cell with fully relaxed lattice parameters.
The band structure for the interpolated lattice parameters is shown in
Fig. \ref{bands}. The in-plane plasma frequency and density of states
and optical conductivities are shown along with those of the
other compounds. As seen, its properties generally interpolate between
those of BaSnO$_3$ and Ba$_3$Sn$_2$O$_7$.

\begin{figure}
\includegraphics[width=\columnwidth,angle=0]{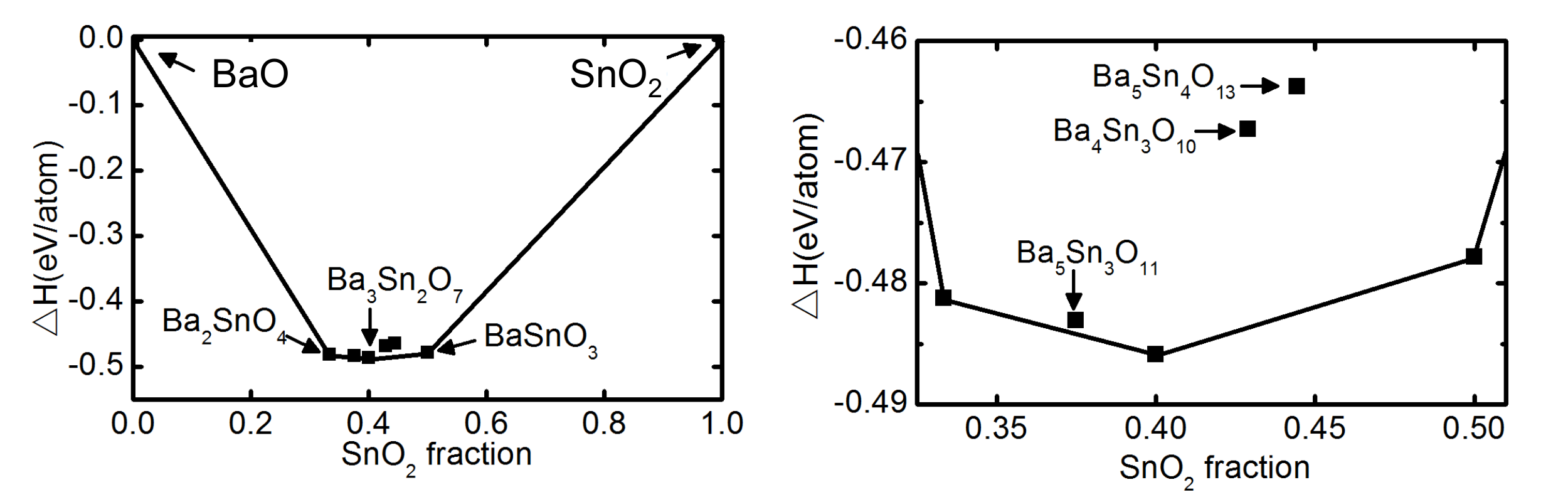}
\caption{Convex hull for the pseudobinary system BaO--SnO$_2$.
The full composition range, with the stable compounds indicated is in the
left panel.
The right panel shows an expanded view of the region studied, with the
compounds above the hull indicated.}
\label{hull}
\end{figure}

We calculated the energetics of the known compounds, along with those of
hypothetical compounds based on addition of an extra BaO
layer between the perovskite blocks in the series, {\em i.e.}
Ba$_5$Sn$_3$O$_{11}$, and the next member of the series, Ba$_5$Sn$_4$O$_{13}$.
The resulting convex hull for the BaO-SnO$_2$ pseudobinary is shown in
Fig. \ref{hull}.
These calculations were done using the PBE-GGA with fully relaxed structures,
to get a consistent set of energies.
As seen, there is a strong tendency towards compound formation in this
pseudobinary. The stable compounds are BaSnO$_3$, Ba$_2$SnO$_4$ and
Ba$_3$Sn$_2$O$_7$.
The other compounds
are above the convex hull, meaning that they are not predicted to be
ground state phases in this pseudobinary. However, they are very close
to the hull, within 0.02 eV/atom
($\sim$ 240 K).
This is certainly below the uncertainty of our density
functional calculations.
Therefore what can be concluded is that these phases are either on
or slightly above the convex hull.
Based on these results it is likely that they
can be stabilized in thin films where kinetic constraints may prevent
phase separation, \cite{hornbostel}
and perhaps also under other conditions.


Experimental results have shown that heavily $n$-type
doped BaSnO$_3$ is promising as a high performance TCO
with good conductivity and optical transparency in the visible.
Layered perovskite variants of this material can be readily
made, and it is likely that they can be layered with the
cubic perovskite material based on the similar chemistry and in-plane
lattice parameter.

In fact, it may well be that conventionally made
films incorporate some fraction of these phases.
The present calculations show that these phases would not necessarily be
harmful for the DC conductivity especially if a high degree of texture
is present.
However, we find that
the presence of doped Ba$_2$SnO$_4$ would be harmful to the
visible light transparency. It is possible in fact that some of the
variable results seen in experiments are due to the presence of 
different amounts of Ba$_2$SnO$_4$ related to
various amounts of Sn deficiency.
It would be desirable to minimize the occurrence of this phase
to improve the transparency.
On the other hand,
the occurrence of bi-layer Ba$_3$Sn$_2$O$_7$
and higher members of the series such as Ba$_4$Sn$_3$O$_{10}$
may not be
detrimental and may add useful functionality, such
as energy selective filtering.
We note that a related concept was recently developed and exploited to produce
low loss, room temperature tunable dielectrics based on SrTiO$_3$,
specifically using the Sr$_{n+1}$Ti$_n$O$_{3n+1}$ 
Ruddlesden-Popper phases with $n$$\geq$3. \cite{lee}
The present results indicate that
it will be interesting to experimentally explore the incorporation of
Ruddlesden-Popper phases into n-type Ba-Sn-O TCO films as a functional
tuning parameter.


YL and YM acknowledge funding support from the Natural Science Foundation
of China (Grant No. 11025418).
LZ acknowledges funding from the Recruitment Program of Global Experts
(the Thousand Young Talents Plan).
Work at ORNL was supported by the Department of Energy, Office of Science,
Basic Energy Sciences,
Materials Sciences and Engineering Division.


\begin{thebibliography}{49}%
\makeatletter
\providecommand \@ifxundefined [1]{%
 \@ifx{#1\undefined}
}%
\providecommand \@ifnum [1]{%
 \ifnum #1\expandafter \@firstoftwo
 \else \expandafter \@secondoftwo
 \fi
}%
\providecommand \@ifx [1]{%
 \ifx #1\expandafter \@firstoftwo
 \else \expandafter \@secondoftwo
 \fi
}%
\providecommand \natexlab [1]{#1}%
\providecommand \enquote  [1]{``#1''}%
\providecommand \bibnamefont  [1]{#1}%
\providecommand \bibfnamefont [1]{#1}%
\providecommand \citenamefont [1]{#1}%
\providecommand \href@noop [0]{\@secondoftwo}%
\providecommand \href [0]{\begingroup \@sanitize@url \@href}%
\providecommand \@href[1]{\@@startlink{#1}\@@href}%
\providecommand \@@href[1]{\endgroup#1\@@endlink}%
\providecommand \@sanitize@url [0]{\catcode `\\12\catcode `\$12\catcode
  `\&12\catcode `\#12\catcode `\^12\catcode `\_12\catcode `\%12\relax}%
\providecommand \@@startlink[1]{}%
\providecommand \@@endlink[0]{}%
\providecommand \url  [0]{\begingroup\@sanitize@url \@url }%
\providecommand \@url [1]{\endgroup\@href {#1}{\urlprefix }}%
\providecommand \urlprefix  [0]{URL }%
\providecommand \Eprint [0]{\href }%
\providecommand \doibase [0]{http://dx.doi.org/}%
\providecommand \selectlanguage [0]{\@gobble}%
\providecommand \bibinfo  [0]{\@secondoftwo}%
\providecommand \bibfield  [0]{\@secondoftwo}%
\providecommand \translation [1]{[#1]}%
\providecommand \BibitemOpen [0]{}%
\providecommand \bibitemStop [0]{}%
\providecommand \bibitemNoStop [0]{.\EOS\space}%
\providecommand \EOS [0]{\spacefactor3000\relax}%
\providecommand \BibitemShut  [1]{\csname bibitem#1\endcsname}%
\let\auto@bib@innerbib\@empty
\bibitem [{\citenamefont {Ginley}\ and\ \citenamefont {Bright}(2000)}]{ginley}%
  \BibitemOpen
  \bibfield  {author} {\bibinfo {author} {\bibfnamefont {D.~S.}\ \bibnamefont
  {Ginley}}\ and\ \bibinfo {author} {\bibfnamefont {C.}~\bibnamefont
  {Bright}},\ }\href@noop {} {\bibfield  {journal} {\bibinfo  {journal} {MRS
  Bull.}\ }\textbf {\bibinfo {volume} {25}},\ \bibinfo {pages} {15} (\bibinfo
  {year} {2000})}\BibitemShut {NoStop}%
\bibitem [{\citenamefont {Kawazoe}\ \emph {et~al.}(2000)\citenamefont
  {Kawazoe}, \citenamefont {Yanagi}, \citenamefont {Ueda},\ and\ \citenamefont
  {Hosono}}]{kawazoe}%
  \BibitemOpen
  \bibfield  {author} {\bibinfo {author} {\bibfnamefont {H.}~\bibnamefont
  {Kawazoe}}, \bibinfo {author} {\bibfnamefont {H.}~\bibnamefont {Yanagi}},
  \bibinfo {author} {\bibfnamefont {K.}~\bibnamefont {Ueda}}, \ and\ \bibinfo
  {author} {\bibfnamefont {H.}~\bibnamefont {Hosono}},\ }\href@noop {}
  {\bibfield  {journal} {\bibinfo  {journal} {MRS Bull.}\ }\textbf {\bibinfo
  {volume} {25}},\ \bibinfo {pages} {28} (\bibinfo {year} {2000})}\BibitemShut
  {NoStop}%
\bibitem [{\citenamefont {Minami}(2008)}]{minami}%
  \BibitemOpen
  \bibfield  {author} {\bibinfo {author} {\bibfnamefont {T.}~\bibnamefont
  {Minami}},\ }\href@noop {} {\bibfield  {journal} {\bibinfo  {journal} {Thin
  Solid Films}\ }\textbf {\bibinfo {volume} {516}},\ \bibinfo {pages} {5822}
  (\bibinfo {year} {2008})}\BibitemShut {NoStop}%
\bibitem [{\citenamefont {Cava}\ \emph {et~al.}(1990)\citenamefont {Cava},
  \citenamefont {Gammel}, \citenamefont {Batlogg}, \citenamefont {Krajewski},\
  and\ \citenamefont {{Peck, Jr.}}}]{cava}%
  \BibitemOpen
  \bibfield  {author} {\bibinfo {author} {\bibfnamefont {R.~J.}\ \bibnamefont
  {Cava}}, \bibinfo {author} {\bibfnamefont {P.}~\bibnamefont {Gammel}},
  \bibinfo {author} {\bibfnamefont {B.}~\bibnamefont {Batlogg}}, \bibinfo
  {author} {\bibfnamefont {J.~J.}\ \bibnamefont {Krajewski}}, \ and\ \bibinfo
  {author} {\bibfnamefont {W.~F.}\ \bibnamefont {{Peck, Jr.}}},\ }\href@noop {}
  {\bibfield  {journal} {\bibinfo  {journal} {Phys. Rev. B}\ }\textbf {\bibinfo
  {volume} {42}},\ \bibinfo {pages} {4815} (\bibinfo {year}
  {1990})}\BibitemShut {NoStop}%
\bibitem [{\citenamefont {Singh}\ \emph {et~al.}(1991)\citenamefont {Singh},
  \citenamefont {Papaconstantopoulos}, \citenamefont {Julien},\ and\
  \citenamefont {{Cyrot-Lackmann}}}]{singh-basno3}%
  \BibitemOpen
  \bibfield  {author} {\bibinfo {author} {\bibfnamefont {D.~J.}\ \bibnamefont
  {Singh}}, \bibinfo {author} {\bibfnamefont {D.~A.}\ \bibnamefont
  {Papaconstantopoulos}}, \bibinfo {author} {\bibfnamefont {J.~P.}\
  \bibnamefont {Julien}}, \ and\ \bibinfo {author} {\bibfnamefont
  {F.}~\bibnamefont {{Cyrot-Lackmann}}},\ }\href@noop {} {\bibfield  {journal}
  {\bibinfo  {journal} {Phys. Rev. B}\ }\textbf {\bibinfo {volume} {44}},\
  \bibinfo {pages} {9519} (\bibinfo {year} {1991})}\BibitemShut {NoStop}%
\bibitem [{\citenamefont {Upadhyay}\ \emph {et~al.}(2004)\citenamefont
  {Upadhyay}, \citenamefont {Parkash},\ and\ \citenamefont {Kumar}}]{upadhyay}%
  \BibitemOpen
  \bibfield  {author} {\bibinfo {author} {\bibfnamefont {S.}~\bibnamefont
  {Upadhyay}}, \bibinfo {author} {\bibfnamefont {O.}~\bibnamefont {Parkash}}, \
  and\ \bibinfo {author} {\bibfnamefont {D.}~\bibnamefont {Kumar}},\
  }\href@noop {} {\bibfield  {journal} {\bibinfo  {journal} {J. Phys. D: Appl.
  Phys.}\ }\textbf {\bibinfo {volume} {37}},\ \bibinfo {pages} {1483} (\bibinfo
  {year} {2004})}\BibitemShut {NoStop}%
\bibitem [{\citenamefont {Hadjarab}\ \emph {et~al.}(2007)\citenamefont
  {Hadjarab}, \citenamefont {Bouguelia},\ and\ \citenamefont
  {Trari}}]{hadjarab}%
  \BibitemOpen
  \bibfield  {author} {\bibinfo {author} {\bibfnamefont {B.}~\bibnamefont
  {Hadjarab}}, \bibinfo {author} {\bibfnamefont {A.}~\bibnamefont {Bouguelia}},
  \ and\ \bibinfo {author} {\bibfnamefont {M.}~\bibnamefont {Trari}},\
  }\href@noop {} {\bibfield  {journal} {\bibinfo  {journal} {J. Phys. D: Appl.
  Phys.}\ }\textbf {\bibinfo {volume} {40}},\ \bibinfo {pages} {5833} (\bibinfo
  {year} {2007})}\BibitemShut {NoStop}%
\bibitem [{\citenamefont {Wang}\ \emph {et~al.}(2007)\citenamefont {Wang},
  \citenamefont {Liu}, \citenamefont {Chen}, \citenamefont {Gao}, \citenamefont
  {Wu},\ and\ \citenamefont {Chen}}]{wang-basno3}%
  \BibitemOpen
  \bibfield  {author} {\bibinfo {author} {\bibfnamefont {H.~F.}\ \bibnamefont
  {Wang}}, \bibinfo {author} {\bibfnamefont {Q.~Z.}\ \bibnamefont {Liu}},
  \bibinfo {author} {\bibfnamefont {F.}~\bibnamefont {Chen}}, \bibinfo {author}
  {\bibfnamefont {G.~Y.}\ \bibnamefont {Gao}}, \bibinfo {author} {\bibfnamefont
  {W.}~\bibnamefont {Wu}}, \ and\ \bibinfo {author} {\bibfnamefont {X.~H.}\
  \bibnamefont {Chen}},\ }\href@noop {} {\bibfield  {journal} {\bibinfo
  {journal} {J. Appl. Phys.}\ }\textbf {\bibinfo {volume} {101}},\ \bibinfo
  {pages} {106105} (\bibinfo {year} {2007})}\BibitemShut {NoStop}%
\bibitem [{\citenamefont {Kim}\ \emph {et~al.}(2012{\natexlab{a}})\citenamefont
  {Kim}, \citenamefont {Kim}, \citenamefont {Kim}, \citenamefont {Kim},
  \citenamefont {Mun}, \citenamefont {Jeon}, \citenamefont {Hong},
  \citenamefont {Lee}, \citenamefont {Ju}, \citenamefont {Kim},\ and\
  \citenamefont {Char}}]{kim}%
  \BibitemOpen
  \bibfield  {author} {\bibinfo {author} {\bibfnamefont {H.~J.}\ \bibnamefont
  {Kim}}, \bibinfo {author} {\bibfnamefont {U.}~\bibnamefont {Kim}}, \bibinfo
  {author} {\bibfnamefont {H.~M.}\ \bibnamefont {Kim}}, \bibinfo {author}
  {\bibfnamefont {T.~H.}\ \bibnamefont {Kim}}, \bibinfo {author} {\bibfnamefont
  {H.~S.}\ \bibnamefont {Mun}}, \bibinfo {author} {\bibfnamefont {B.~G.}\
  \bibnamefont {Jeon}}, \bibinfo {author} {\bibfnamefont {K.~T.}\ \bibnamefont
  {Hong}}, \bibinfo {author} {\bibfnamefont {W.~J.}\ \bibnamefont {Lee}},
  \bibinfo {author} {\bibfnamefont {C.}~\bibnamefont {Ju}}, \bibinfo {author}
  {\bibfnamefont {K.~H.}\ \bibnamefont {Kim}}, \ and\ \bibinfo {author}
  {\bibfnamefont {K.}~\bibnamefont {Char}},\ }\href@noop {} {\bibfield
  {journal} {\bibinfo  {journal} {Appl. Phys. Express}\ }\textbf {\bibinfo
  {volume} {5}},\ \bibinfo {pages} {061102} (\bibinfo {year}
  {2012}{\natexlab{a}})}\BibitemShut {NoStop}%
\bibitem [{\citenamefont {Luo}\ \emph {et~al.}(2012)\citenamefont {Luo},
  \citenamefont {Oh}, \citenamefont {Sirenko}, \citenamefont {Gao},
  \citenamefont {Tyson}, \citenamefont {Char},\ and\ \citenamefont
  {Cheong}}]{luo}%
  \BibitemOpen
  \bibfield  {author} {\bibinfo {author} {\bibfnamefont {X.}~\bibnamefont
  {Luo}}, \bibinfo {author} {\bibfnamefont {Y.~S.}\ \bibnamefont {Oh}},
  \bibinfo {author} {\bibfnamefont {A.}~\bibnamefont {Sirenko}}, \bibinfo
  {author} {\bibfnamefont {P.}~\bibnamefont {Gao}}, \bibinfo {author}
  {\bibfnamefont {T.~A.}\ \bibnamefont {Tyson}}, \bibinfo {author}
  {\bibfnamefont {K.}~\bibnamefont {Char}}, \ and\ \bibinfo {author}
  {\bibfnamefont {S.~W.}\ \bibnamefont {Cheong}},\ }\href@noop {} {\bibfield
  {journal} {\bibinfo  {journal} {Appl. Phys. Lett.}\ }\textbf {\bibinfo
  {volume} {100}},\ \bibinfo {pages} {172112} (\bibinfo {year}
  {2012})}\BibitemShut {NoStop}%
\bibitem [{\citenamefont {Liu}\ \emph {et~al.}(2012)\citenamefont {Liu},
  \citenamefont {Liu}, \citenamefont {Li}, \citenamefont {Li}, \citenamefont
  {Zhu}, \citenamefont {Dai}, \citenamefont {Liu}, \citenamefont {Zhang},\ and\
  \citenamefont {Dai}}]{liu2}%
  \BibitemOpen
  \bibfield  {author} {\bibinfo {author} {\bibfnamefont {Q.}~\bibnamefont
  {Liu}}, \bibinfo {author} {\bibfnamefont {J.}~\bibnamefont {Liu}}, \bibinfo
  {author} {\bibfnamefont {B.}~\bibnamefont {Li}}, \bibinfo {author}
  {\bibfnamefont {H.}~\bibnamefont {Li}}, \bibinfo {author} {\bibfnamefont
  {G.}~\bibnamefont {Zhu}}, \bibinfo {author} {\bibfnamefont {K.}~\bibnamefont
  {Dai}}, \bibinfo {author} {\bibfnamefont {Z.}~\bibnamefont {Liu}}, \bibinfo
  {author} {\bibfnamefont {P.}~\bibnamefont {Zhang}}, \ and\ \bibinfo {author}
  {\bibfnamefont {J.}~\bibnamefont {Dai}},\ }\href@noop {} {\bibfield
  {journal} {\bibinfo  {journal} {Appl. Phys. Lett.}\ }\textbf {\bibinfo
  {volume} {101}},\ \bibinfo {pages} {241901} (\bibinfo {year}
  {2012})}\BibitemShut {NoStop}%
\bibitem [{\citenamefont {Kim}\ \emph {et~al.}(2012{\natexlab{b}})\citenamefont
  {Kim}, \citenamefont {Kim}, \citenamefont {Kim}, \citenamefont {Kim},
  \citenamefont {Kim}, \citenamefont {Jeon}, \citenamefont {Lee}, \citenamefont
  {Mun}, \citenamefont {Hong}, \citenamefont {Yu}, \citenamefont {Char},\ and\
  \citenamefont {Kim}}]{kim2}%
  \BibitemOpen
  \bibfield  {author} {\bibinfo {author} {\bibfnamefont {H.~J.}\ \bibnamefont
  {Kim}}, \bibinfo {author} {\bibfnamefont {U.}~\bibnamefont {Kim}}, \bibinfo
  {author} {\bibfnamefont {T.~H.}\ \bibnamefont {Kim}}, \bibinfo {author}
  {\bibfnamefont {J.}~\bibnamefont {Kim}}, \bibinfo {author} {\bibfnamefont
  {H.~M.}\ \bibnamefont {Kim}}, \bibinfo {author} {\bibfnamefont {B.-G.}\
  \bibnamefont {Jeon}}, \bibinfo {author} {\bibfnamefont {W.-J.}\ \bibnamefont
  {Lee}}, \bibinfo {author} {\bibfnamefont {H.~S.}\ \bibnamefont {Mun}},
  \bibinfo {author} {\bibfnamefont {K.~T.}\ \bibnamefont {Hong}}, \bibinfo
  {author} {\bibfnamefont {J.}~\bibnamefont {Yu}}, \bibinfo {author}
  {\bibfnamefont {K.}~\bibnamefont {Char}}, \ and\ \bibinfo {author}
  {\bibfnamefont {K.~H.}\ \bibnamefont {Kim}},\ }\href@noop {} {\bibfield
  {journal} {\bibinfo  {journal} {Phys. Rev. B}\ }\textbf {\bibinfo {volume}
  {86}},\ \bibinfo {pages} {165205} (\bibinfo {year}
  {2012}{\natexlab{b}})}\BibitemShut {NoStop}%
\bibitem [{\citenamefont {Mizoguchi}\ \emph {et~al.}(2004)\citenamefont
  {Mizoguchi}, \citenamefont {Eng},\ and\ \citenamefont
  {Woodward}}]{mizoguchi}%
  \BibitemOpen
  \bibfield  {author} {\bibinfo {author} {\bibfnamefont {H.}~\bibnamefont
  {Mizoguchi}}, \bibinfo {author} {\bibfnamefont {H.~W.}\ \bibnamefont {Eng}},
  \ and\ \bibinfo {author} {\bibfnamefont {P.~M.}\ \bibnamefont {Woodward}},\
  }\href@noop {} {\bibfield  {journal} {\bibinfo  {journal} {Inorg. Chem.}\
  }\textbf {\bibinfo {volume} {43}},\ \bibinfo {pages} {1667} (\bibinfo {year}
  {2004})}\BibitemShut {NoStop}%
\bibitem [{\citenamefont {Mizoguchi}\ \emph {et~al.}(2013)\citenamefont
  {Mizoguchi}, \citenamefont {Chen}, \citenamefont {Boolchand}, \citenamefont
  {Ksenofontov}, \citenamefont {Felser}, \citenamefont {Barnes},\ and\
  \citenamefont {Woodward}}]{mizoguchi-ba}%
  \BibitemOpen
  \bibfield  {author} {\bibinfo {author} {\bibfnamefont {H.}~\bibnamefont
  {Mizoguchi}}, \bibinfo {author} {\bibfnamefont {P.}~\bibnamefont {Chen}},
  \bibinfo {author} {\bibfnamefont {P.}~\bibnamefont {Boolchand}}, \bibinfo
  {author} {\bibfnamefont {V.}~\bibnamefont {Ksenofontov}}, \bibinfo {author}
  {\bibfnamefont {C.}~\bibnamefont {Felser}}, \bibinfo {author} {\bibfnamefont
  {P.~W.}\ \bibnamefont {Barnes}}, \ and\ \bibinfo {author} {\bibfnamefont
  {P.~M.}\ \bibnamefont {Woodward}},\ }\href@noop {} {\bibfield  {journal}
  {\bibinfo  {journal} {Chem. Mater.}\ }\textbf {\bibinfo {volume} {25}},\
  \bibinfo {pages} {3858} (\bibinfo {year} {2013})}\BibitemShut {NoStop}%
\bibitem [{\citenamefont {Fan}\ \emph {et~al.}(2014)\citenamefont {Fan},
  \citenamefont {Zheng}, \citenamefont {Chen},\ and\ \citenamefont
  {Singh}}]{fan}%
  \BibitemOpen
  \bibfield  {author} {\bibinfo {author} {\bibfnamefont {X.~F.}\ \bibnamefont
  {Fan}}, \bibinfo {author} {\bibfnamefont {W.~T.}\ \bibnamefont {Zheng}},
  \bibinfo {author} {\bibfnamefont {X.}~\bibnamefont {Chen}}, \ and\ \bibinfo
  {author} {\bibfnamefont {D.~J.}\ \bibnamefont {Singh}},\ }\href@noop {}
  {\bibfield  {journal} {\bibinfo  {journal} {PLoS One}\ }\textbf {\bibinfo
  {volume} {9}},\ \bibinfo {pages} {e91423} (\bibinfo {year}
  {2014})}\BibitemShut {NoStop}%
\bibitem [{\citenamefont {Liu}\ \emph {et~al.}(2013)\citenamefont {Liu},
  \citenamefont {Yang}, \citenamefont {Xiang},\ and\ \citenamefont
  {Wei}}]{liu}%
  \BibitemOpen
  \bibfield  {author} {\bibinfo {author} {\bibfnamefont {H.~R.}\ \bibnamefont
  {Liu}}, \bibinfo {author} {\bibfnamefont {J.~H.}\ \bibnamefont {Yang}},
  \bibinfo {author} {\bibfnamefont {H.~J.}\ \bibnamefont {Xiang}}, \ and\
  \bibinfo {author} {\bibfnamefont {S.~H.}\ \bibnamefont {Wei}},\ }\href@noop
  {} {\bibfield  {journal} {\bibinfo  {journal} {Appl. Phys. Lett.}\ }\textbf
  {\bibinfo {volume} {102}},\ \bibinfo {pages} {112109} (\bibinfo {year}
  {2013})}\BibitemShut {NoStop}%
\bibitem [{\citenamefont {Singh}\ \emph {et~al.}(2014)\citenamefont {Singh},
  \citenamefont {Xu},\ and\ \citenamefont {Ong}}]{singh3}%
  \BibitemOpen
  \bibfield  {author} {\bibinfo {author} {\bibfnamefont {D.~J.}\ \bibnamefont
  {Singh}}, \bibinfo {author} {\bibfnamefont {Q.}~\bibnamefont {Xu}}, \ and\
  \bibinfo {author} {\bibfnamefont {K.~P.}\ \bibnamefont {Ong}},\ }\href@noop
  {} {\bibfield  {journal} {\bibinfo  {journal} {Appl. Phys. Lett.}\ }\textbf
  {\bibinfo {volume} {104}},\ \bibinfo {pages} {011910} (\bibinfo {year}
  {2014})}\BibitemShut {NoStop}%
\bibitem [{\citenamefont {Mun}\ \emph {et~al.}(2013)\citenamefont {Mun},
  \citenamefont {Kim}, \citenamefont {Kim}, \citenamefont {Park}, \citenamefont
  {Kim}, \citenamefont {Kim}, \citenamefont {Kim},\ and\ \citenamefont
  {Char}}]{mun}%
  \BibitemOpen
  \bibfield  {author} {\bibinfo {author} {\bibfnamefont {H.}~\bibnamefont
  {Mun}}, \bibinfo {author} {\bibfnamefont {U.}~\bibnamefont {Kim}}, \bibinfo
  {author} {\bibfnamefont {H.~M.}\ \bibnamefont {Kim}}, \bibinfo {author}
  {\bibfnamefont {C.}~\bibnamefont {Park}}, \bibinfo {author} {\bibfnamefont
  {T.~H.}\ \bibnamefont {Kim}}, \bibinfo {author} {\bibfnamefont {H.~J.}\
  \bibnamefont {Kim}}, \bibinfo {author} {\bibfnamefont {K.~H.}\ \bibnamefont
  {Kim}}, \ and\ \bibinfo {author} {\bibfnamefont {K.}~\bibnamefont {Char}},\
  }\href@noop {} {\bibfield  {journal} {\bibinfo  {journal} {Appl. Phys.
  Lett.}\ }\textbf {\bibinfo {volume} {102}},\ \bibinfo {pages} {252105}
  (\bibinfo {year} {2013})}\BibitemShut {NoStop}%
\bibitem [{\citenamefont {Wadekar}\ \emph {et~al.}(2014)\citenamefont
  {Wadekar}, \citenamefont {Alaria}, \citenamefont {{O'Sullivan}},
  \citenamefont {Flack}, \citenamefont {Manning}, \citenamefont {Phillips},
  \citenamefont {Durose}, \citenamefont {Lozano}, \citenamefont {Lucas},
  \citenamefont {Claridge},\ and\ \citenamefont {Rosseinsky}}]{wadekar}%
  \BibitemOpen
  \bibfield  {author} {\bibinfo {author} {\bibfnamefont {P.~V.}\ \bibnamefont
  {Wadekar}}, \bibinfo {author} {\bibfnamefont {J.}~\bibnamefont {Alaria}},
  \bibinfo {author} {\bibfnamefont {M.}~\bibnamefont {{O'Sullivan}}}, \bibinfo
  {author} {\bibfnamefont {N.~L.~O.}\ \bibnamefont {Flack}}, \bibinfo {author}
  {\bibfnamefont {T.~D.}\ \bibnamefont {Manning}}, \bibinfo {author}
  {\bibfnamefont {L.~J.}\ \bibnamefont {Phillips}}, \bibinfo {author}
  {\bibfnamefont {K.}~\bibnamefont {Durose}}, \bibinfo {author} {\bibfnamefont
  {O.}~\bibnamefont {Lozano}}, \bibinfo {author} {\bibfnamefont
  {S.}~\bibnamefont {Lucas}}, \bibinfo {author} {\bibfnamefont {J.~B.}\
  \bibnamefont {Claridge}}, \ and\ \bibinfo {author} {\bibfnamefont {M.~J.}\
  \bibnamefont {Rosseinsky}},\ }\href@noop {} {\bibfield  {journal} {\bibinfo
  {journal} {Appl. Phys. Lett.}\ }\textbf {\bibinfo {volume} {105}},\ \bibinfo
  {pages} {052104} (\bibinfo {year} {2014})}\BibitemShut {NoStop}%
\bibitem [{\citenamefont {Sallis}\ \emph {et~al.}(2013)\citenamefont {Sallis},
  \citenamefont {Scanlon}, \citenamefont {Chae}, \citenamefont {Quackenbush},
  \citenamefont {Fischer}, \citenamefont {Woicik}, \citenamefont {Guo},
  \citenamefont {Cheong},\ and\ \citenamefont {Piper}}]{sallis}%
  \BibitemOpen
  \bibfield  {author} {\bibinfo {author} {\bibfnamefont {S.}~\bibnamefont
  {Sallis}}, \bibinfo {author} {\bibfnamefont {D.~O.}\ \bibnamefont {Scanlon}},
  \bibinfo {author} {\bibfnamefont {S.~C.}\ \bibnamefont {Chae}}, \bibinfo
  {author} {\bibfnamefont {N.~F.}\ \bibnamefont {Quackenbush}}, \bibinfo
  {author} {\bibfnamefont {D.~A.}\ \bibnamefont {Fischer}}, \bibinfo {author}
  {\bibfnamefont {J.~C.}\ \bibnamefont {Woicik}}, \bibinfo {author}
  {\bibfnamefont {J.~H.}\ \bibnamefont {Guo}}, \bibinfo {author} {\bibfnamefont
  {S.~W.}\ \bibnamefont {Cheong}}, \ and\ \bibinfo {author} {\bibfnamefont
  {L.~F.~J.}\ \bibnamefont {Piper}},\ }\href@noop {} {\bibfield  {journal}
  {\bibinfo  {journal} {Appl. Phys. Lett.}\ }\textbf {\bibinfo {volume}
  {103}},\ \bibinfo {pages} {042105} (\bibinfo {year} {2013})}\BibitemShut
  {NoStop}%
\bibitem [{\citenamefont {Scanlon}(2013)}]{scanlon}%
  \BibitemOpen
  \bibfield  {author} {\bibinfo {author} {\bibfnamefont {D.~O.}\ \bibnamefont
  {Scanlon}},\ }\href@noop {} {\bibfield  {journal} {\bibinfo  {journal} {Phys.
  Rev. B}\ }\textbf {\bibinfo {volume} {87}},\ \bibinfo {pages} {161201}
  (\bibinfo {year} {2013})}\BibitemShut {NoStop}%
\bibitem [{\citenamefont {Kim}\ \emph {et~al.}(2013)\citenamefont {Kim},
  \citenamefont {Jo},\ and\ \citenamefont {Cheong}}]{kim3}%
  \BibitemOpen
  \bibfield  {author} {\bibinfo {author} {\bibfnamefont {B.~G.}\ \bibnamefont
  {Kim}}, \bibinfo {author} {\bibfnamefont {J.~Y.}\ \bibnamefont {Jo}}, \ and\
  \bibinfo {author} {\bibfnamefont {S.~W.}\ \bibnamefont {Cheong}},\
  }\href@noop {} {\bibfield  {journal} {\bibinfo  {journal} {J. Solid State
  Chem.}\ }\textbf {\bibinfo {volume} {197}},\ \bibinfo {pages} {134} (\bibinfo
  {year} {2013})}\BibitemShut {NoStop}%
\bibitem [{\citenamefont {Wagner}\ and\ \citenamefont {Binder}(1959)}]{wagner}%
  \BibitemOpen
  \bibfield  {author} {\bibinfo {author} {\bibfnamefont {G.}~\bibnamefont
  {Wagner}}\ and\ \bibinfo {author} {\bibfnamefont {H.}~\bibnamefont
  {Binder}},\ }\href@noop {} {\bibfield  {journal} {\bibinfo  {journal} {Z.
  Anorg. Allg. Chem.}\ }\textbf {\bibinfo {volume} {298}},\ \bibinfo {pages}
  {12} (\bibinfo {year} {1959})}\BibitemShut {NoStop}%
\bibitem [{\citenamefont {Hinatsu}\ and\ \citenamefont
  {Tezuka}(1998)}]{hinatsu}%
  \BibitemOpen
  \bibfield  {author} {\bibinfo {author} {\bibfnamefont {Y.}~\bibnamefont
  {Hinatsu}}\ and\ \bibinfo {author} {\bibfnamefont {K.}~\bibnamefont
  {Tezuka}},\ }\href@noop {} {\bibfield  {journal} {\bibinfo  {journal} {J.
  Solid State Chem.}\ }\textbf {\bibinfo {volume} {138}},\ \bibinfo {pages}
  {329} (\bibinfo {year} {1998})}\BibitemShut {NoStop}%
\bibitem [{\citenamefont {Green}\ \emph {et~al.}(1996)\citenamefont {Green},
  \citenamefont {Prassides}, \citenamefont {Day},\ and\ \citenamefont
  {Neumann}}]{green}%
  \BibitemOpen
  \bibfield  {author} {\bibinfo {author} {\bibfnamefont {M.~A.}\ \bibnamefont
  {Green}}, \bibinfo {author} {\bibfnamefont {K.}~\bibnamefont {Prassides}},
  \bibinfo {author} {\bibfnamefont {P.}~\bibnamefont {Day}}, \ and\ \bibinfo
  {author} {\bibfnamefont {D.~A.}\ \bibnamefont {Neumann}},\ }\href@noop {}
  {\bibfield  {journal} {\bibinfo  {journal} {J. Chem. Soc., Faraday Trans.}\
  }\textbf {\bibinfo {volume} {92}},\ \bibinfo {pages} {2155} (\bibinfo {year}
  {1996})}\BibitemShut {NoStop}%
\bibitem [{\citenamefont {Kennedy}(1997)}]{kennedy}%
  \BibitemOpen
  \bibfield  {author} {\bibinfo {author} {\bibfnamefont {B.~J.}\ \bibnamefont
  {Kennedy}},\ }\href@noop {} {\bibfield  {journal} {\bibinfo  {journal} {Aust.
  J. Chem.}\ }\textbf {\bibinfo {volume} {50}},\ \bibinfo {pages} {917}
  (\bibinfo {year} {1997})}\BibitemShut {NoStop}%
\bibitem [{\citenamefont {Ropp}(2012)}]{ropp}%
  \BibitemOpen
  \bibfield  {author} {\bibinfo {author} {\bibfnamefont {R.~C.}\ \bibnamefont
  {Ropp}},\ }\href@noop {} {\emph {\bibinfo {title} {{Encyclopedia of the
  Alkaline Earth Compounds}}}}\ (\bibinfo  {publisher} {Elsevier, Amsterdam},\
  \bibinfo {year} {2012})\ p.\ \bibinfo {pages} {454}\BibitemShut {NoStop}%
\bibitem [{\citenamefont {Yamashita}\ and\ \citenamefont
  {Ueda}(2007)}]{yamashita}%
  \BibitemOpen
  \bibfield  {author} {\bibinfo {author} {\bibfnamefont {T.}~\bibnamefont
  {Yamashita}}\ and\ \bibinfo {author} {\bibfnamefont {K.}~\bibnamefont
  {Ueda}},\ }\href@noop {} {\bibfield  {journal} {\bibinfo  {journal} {J. Solid
  State Chem.}\ }\textbf {\bibinfo {volume} {180}},\ \bibinfo {pages} {1410}
  (\bibinfo {year} {2007})}\BibitemShut {NoStop}%
\bibitem [{\citenamefont {Stanulis}\ \emph {et~al.}(2012)\citenamefont
  {Stanulis}, \citenamefont {Sakirzanovas}, \citenamefont {{Van Bael}},\ and\
  \citenamefont {Kareiva}}]{stanulis}%
  \BibitemOpen
  \bibfield  {author} {\bibinfo {author} {\bibfnamefont {A.}~\bibnamefont
  {Stanulis}}, \bibinfo {author} {\bibfnamefont {S.}~\bibnamefont
  {Sakirzanovas}}, \bibinfo {author} {\bibfnamefont {M.}~\bibnamefont {{Van
  Bael}}}, \ and\ \bibinfo {author} {\bibfnamefont {A.}~\bibnamefont
  {Kareiva}},\ }\href@noop {} {\bibfield  {journal} {\bibinfo  {journal} {J.
  Sol-Gel Sci. Technol.}\ }\textbf {\bibinfo {volume} {64}},\ \bibinfo {pages}
  {643} (\bibinfo {year} {2012})}\BibitemShut {NoStop}%
\bibitem [{\citenamefont {Lee}\ \emph {et~al.}(2013)\citenamefont {Lee},
  \citenamefont {Orloff}, \citenamefont {Biroi}, \citenamefont {Zhu},
  \citenamefont {Goian}, \citenamefont {Rocas}, \citenamefont {Haislmaier},
  \citenamefont {Vlahos}, \citenamefont {Mundy}, \citenamefont {Kourkoutis},
  \citenamefont {Nie}, \citenamefont {Biegalski}, \citenamefont {Zhang},
  \citenamefont {Bernhagen}, \citenamefont {Benedek}, \citenamefont {Kim},
  \citenamefont {Brock}, \citenamefont {Uecker}, \citenamefont {Xi},
  \citenamefont {Gopalan}, \citenamefont {Nuzhnyy}, \citenamefont {Kamba},
  \citenamefont {Muller}, \citenamefont {Takeuchi}, \citenamefont {Booth},
  \citenamefont {Fennie},\ and\ \citenamefont {Schlom}}]{lee}%
  \BibitemOpen
  \bibfield  {author} {\bibinfo {author} {\bibfnamefont {C.~H.}\ \bibnamefont
  {Lee}}, \bibinfo {author} {\bibfnamefont {N.~D.}\ \bibnamefont {Orloff}},
  \bibinfo {author} {\bibfnamefont {T.}~\bibnamefont {Biroi}}, \bibinfo
  {author} {\bibfnamefont {Y.}~\bibnamefont {Zhu}}, \bibinfo {author}
  {\bibfnamefont {V.}~\bibnamefont {Goian}}, \bibinfo {author} {\bibfnamefont
  {E.}~\bibnamefont {Rocas}}, \bibinfo {author} {\bibfnamefont
  {R.}~\bibnamefont {Haislmaier}}, \bibinfo {author} {\bibfnamefont
  {E.}~\bibnamefont {Vlahos}}, \bibinfo {author} {\bibfnamefont {J.~A.}\
  \bibnamefont {Mundy}}, \bibinfo {author} {\bibfnamefont {L.~F.}\ \bibnamefont
  {Kourkoutis}}, \bibinfo {author} {\bibfnamefont {Y.}~\bibnamefont {Nie}},
  \bibinfo {author} {\bibfnamefont {M.~D.}\ \bibnamefont {Biegalski}}, \bibinfo
  {author} {\bibfnamefont {J.}~\bibnamefont {Zhang}}, \bibinfo {author}
  {\bibfnamefont {M.}~\bibnamefont {Bernhagen}}, \bibinfo {author}
  {\bibfnamefont {N.~A.}\ \bibnamefont {Benedek}}, \bibinfo {author}
  {\bibfnamefont {Y.}~\bibnamefont {Kim}}, \bibinfo {author} {\bibfnamefont
  {J.~D.}\ \bibnamefont {Brock}}, \bibinfo {author} {\bibfnamefont
  {R.}~\bibnamefont {Uecker}}, \bibinfo {author} {\bibfnamefont {X.~X.}\
  \bibnamefont {Xi}}, \bibinfo {author} {\bibfnamefont {V.}~\bibnamefont
  {Gopalan}}, \bibinfo {author} {\bibfnamefont {D.}~\bibnamefont {Nuzhnyy}},
  \bibinfo {author} {\bibfnamefont {S.}~\bibnamefont {Kamba}}, \bibinfo
  {author} {\bibfnamefont {D.~A.}\ \bibnamefont {Muller}}, \bibinfo {author}
  {\bibfnamefont {I.}~\bibnamefont {Takeuchi}}, \bibinfo {author}
  {\bibfnamefont {J.~C.}\ \bibnamefont {Booth}}, \bibinfo {author}
  {\bibfnamefont {C.~J.}\ \bibnamefont {Fennie}}, \ and\ \bibinfo {author}
  {\bibfnamefont {D.~G.}\ \bibnamefont {Schlom}},\ }\href@noop {} {\bibfield
  {journal} {\bibinfo  {journal} {Nature (London)}\ }\textbf {\bibinfo {volume}
  {502}},\ \bibinfo {pages} {532} (\bibinfo {year} {2013})}\BibitemShut
  {NoStop}%
\bibitem [{\citenamefont {Green}\ \emph {et~al.}(2000)\citenamefont {Green},
  \citenamefont {Prassides}, \citenamefont {Day},\ and\ \citenamefont
  {Newmann}}]{green2}%
  \BibitemOpen
  \bibfield  {author} {\bibinfo {author} {\bibfnamefont {M.~A.}\ \bibnamefont
  {Green}}, \bibinfo {author} {\bibfnamefont {K.}~\bibnamefont {Prassides}},
  \bibinfo {author} {\bibfnamefont {P.}~\bibnamefont {Day}}, \ and\ \bibinfo
  {author} {\bibfnamefont {D.~A.}\ \bibnamefont {Newmann}},\ }\href@noop {}
  {\bibfield  {journal} {\bibinfo  {journal} {Int. J. Inorg. Mater.}\ }\textbf
  {\bibinfo {volume} {2}},\ \bibinfo {pages} {35} (\bibinfo {year}
  {2000})}\BibitemShut {NoStop}%
\bibitem [{\citenamefont {Fu}\ \emph {et~al.}(2004)\citenamefont {Fu},
  \citenamefont {Visser}, \citenamefont {Knight},\ and\ \citenamefont
  {Ijdo}}]{fu}%
  \BibitemOpen
  \bibfield  {author} {\bibinfo {author} {\bibfnamefont {W.~T.}\ \bibnamefont
  {Fu}}, \bibinfo {author} {\bibfnamefont {D.}~\bibnamefont {Visser}}, \bibinfo
  {author} {\bibfnamefont {K.~S.}\ \bibnamefont {Knight}}, \ and\ \bibinfo
  {author} {\bibfnamefont {D.~J.~W.}\ \bibnamefont {Ijdo}},\ }\href@noop {}
  {\bibfield  {journal} {\bibinfo  {journal} {J. Solid State Chem.}\ }\textbf
  {\bibinfo {volume} {177}},\ \bibinfo {pages} {4081} (\bibinfo {year}
  {2004})}\BibitemShut {NoStop}%
\bibitem [{\citenamefont {Kamimura}\ \emph {et~al.}(2012)\citenamefont
  {Kamimura}, \citenamefont {Yamada},\ and\ \citenamefont {Xu}}]{kamimura}%
  \BibitemOpen
  \bibfield  {author} {\bibinfo {author} {\bibfnamefont {S.}~\bibnamefont
  {Kamimura}}, \bibinfo {author} {\bibfnamefont {H.}~\bibnamefont {Yamada}}, \
  and\ \bibinfo {author} {\bibfnamefont {C.~N.}\ \bibnamefont {Xu}},\
  }\href@noop {} {\bibfield  {journal} {\bibinfo  {journal} {Appl. Phys.
  Lett.}\ }\textbf {\bibinfo {volume} {101}},\ \bibinfo {pages} {091113}
  (\bibinfo {year} {2012})}\BibitemShut {NoStop}%
\bibitem [{\citenamefont {Vegas}\ \emph {et~al.}(1986)\citenamefont {Vegas},
  \citenamefont {{Vallet-Regi}}, \citenamefont {{Gonzalez-Calbet}},\ and\
  \citenamefont {{Alario-Franco}}}]{vegas}%
  \BibitemOpen
  \bibfield  {author} {\bibinfo {author} {\bibfnamefont {A.}~\bibnamefont
  {Vegas}}, \bibinfo {author} {\bibfnamefont {M.}~\bibnamefont
  {{Vallet-Regi}}}, \bibinfo {author} {\bibfnamefont {J.~M.}\ \bibnamefont
  {{Gonzalez-Calbet}}}, \ and\ \bibinfo {author} {\bibfnamefont {M.~A.}\
  \bibnamefont {{Alario-Franco}}},\ }\href@noop {} {\bibfield  {journal}
  {\bibinfo  {journal} {Acta Cryst. B}\ }\textbf {\bibinfo {volume} {42}},\
  \bibinfo {pages} {167} (\bibinfo {year} {1986})}\BibitemShut {NoStop}%
\bibitem [{\citenamefont {Maekawa}\ \emph {et~al.}(2006)\citenamefont
  {Maekawa}, \citenamefont {Kurosaki},\ and\ \citenamefont
  {Yamanaka}}]{maekawa}%
  \BibitemOpen
  \bibfield  {author} {\bibinfo {author} {\bibfnamefont {T.}~\bibnamefont
  {Maekawa}}, \bibinfo {author} {\bibfnamefont {K.}~\bibnamefont {Kurosaki}}, \
  and\ \bibinfo {author} {\bibfnamefont {S.}~\bibnamefont {Yamanaka}},\
  }\href@noop {} {\bibfield  {journal} {\bibinfo  {journal} {J. Alloys
  Compds.}\ }\textbf {\bibinfo {volume} {416}},\ \bibinfo {pages} {214}
  (\bibinfo {year} {2006})}\BibitemShut {NoStop}%
\bibitem [{\citenamefont {Bevillon}\ \emph {et~al.}(2008)\citenamefont
  {Bevillon}, \citenamefont {Chesnaud}, \citenamefont {Wang}, \citenamefont
  {Dezanneau},\ and\ \citenamefont {Geneste}}]{bevillon}%
  \BibitemOpen
  \bibfield  {author} {\bibinfo {author} {\bibfnamefont {E.}~\bibnamefont
  {Bevillon}}, \bibinfo {author} {\bibfnamefont {A.}~\bibnamefont {Chesnaud}},
  \bibinfo {author} {\bibfnamefont {Y.}~\bibnamefont {Wang}}, \bibinfo {author}
  {\bibfnamefont {G.}~\bibnamefont {Dezanneau}}, \ and\ \bibinfo {author}
  {\bibfnamefont {G.}~\bibnamefont {Geneste}},\ }\href@noop {} {\bibfield
  {journal} {\bibinfo  {journal} {J. Phys.:Condens. Matter}\ }\textbf {\bibinfo
  {volume} {20}},\ \bibinfo {pages} {145217} (\bibinfo {year}
  {2008})}\BibitemShut {NoStop}%
\bibitem [{\citenamefont {Singh}\ and\ \citenamefont
  {Nordstrom}(2006)}]{singh-book}%
  \BibitemOpen
  \bibfield  {author} {\bibinfo {author} {\bibfnamefont {D.~J.}\ \bibnamefont
  {Singh}}\ and\ \bibinfo {author} {\bibfnamefont {L.}~\bibnamefont
  {Nordstrom}},\ }\href@noop {} {\emph {\bibinfo {title} {{Planewaves
  Pseudopotentials and the LAPW Method, 2nd Edition}}}}\ (\bibinfo  {publisher}
  {Springer,Berlin},\ \bibinfo {year} {2006})\BibitemShut {NoStop}%
\bibitem [{\citenamefont {Heyd}\ \emph {et~al.}(2003)\citenamefont {Heyd},
  \citenamefont {Scuseria},\ and\ \citenamefont {Ernzerhof}}]{heyd}%
  \BibitemOpen
  \bibfield  {author} {\bibinfo {author} {\bibfnamefont {J.}~\bibnamefont
  {Heyd}}, \bibinfo {author} {\bibfnamefont {G.~E.}\ \bibnamefont {Scuseria}},
  \ and\ \bibinfo {author} {\bibfnamefont {M.}~\bibnamefont {Ernzerhof}},\
  }\href@noop {} {\bibfield  {journal} {\bibinfo  {journal} {J. Chem. Phys.}\
  }\textbf {\bibinfo {volume} {118}},\ \bibinfo {pages} {8207} (\bibinfo {year}
  {2003})}\BibitemShut {NoStop}%
\bibitem [{\citenamefont {Heyd}\ \emph {et~al.}(2006)\citenamefont {Heyd},
  \citenamefont {Scuseria},\ and\ \citenamefont {Ernzerhof}}]{heyd1}%
  \BibitemOpen
  \bibfield  {author} {\bibinfo {author} {\bibfnamefont {J.}~\bibnamefont
  {Heyd}}, \bibinfo {author} {\bibfnamefont {G.~E.}\ \bibnamefont {Scuseria}},
  \ and\ \bibinfo {author} {\bibfnamefont {M.}~\bibnamefont {Ernzerhof}},\
  }\href@noop {} {\bibfield  {journal} {\bibinfo  {journal} {J. Chem. Phys.}\
  }\textbf {\bibinfo {volume} {124}},\ \bibinfo {pages} {219906} (\bibinfo
  {year} {2006})}\BibitemShut {NoStop}%
\bibitem [{\citenamefont {Perdew}\ \emph {et~al.}(1996)\citenamefont {Perdew},
  \citenamefont {Burke},\ and\ \citenamefont {Ernzerhof}}]{pbe}%
  \BibitemOpen
  \bibfield  {author} {\bibinfo {author} {\bibfnamefont {J.~P.}\ \bibnamefont
  {Perdew}}, \bibinfo {author} {\bibfnamefont {K.}~\bibnamefont {Burke}}, \
  and\ \bibinfo {author} {\bibfnamefont {M.}~\bibnamefont {Ernzerhof}},\
  }\href@noop {} {\bibfield  {journal} {\bibinfo  {journal} {Phys. Rev. Lett.}\
  }\textbf {\bibinfo {volume} {77}},\ \bibinfo {pages} {3865} (\bibinfo {year}
  {1996})}\BibitemShut {NoStop}%
\bibitem [{\citenamefont {Sjostedt}\ \emph {et~al.}(2000)\citenamefont
  {Sjostedt}, \citenamefont {Nordstrom},\ and\ \citenamefont {Singh}}]{sjo}%
  \BibitemOpen
  \bibfield  {author} {\bibinfo {author} {\bibfnamefont {E.}~\bibnamefont
  {Sjostedt}}, \bibinfo {author} {\bibfnamefont {L.}~\bibnamefont {Nordstrom}},
  \ and\ \bibinfo {author} {\bibfnamefont {D.~J.}\ \bibnamefont {Singh}},\
  }\href@noop {} {\bibfield  {journal} {\bibinfo  {journal} {Solid State
  Commun.}\ }\textbf {\bibinfo {volume} {114}},\ \bibinfo {pages} {15}
  (\bibinfo {year} {2000})}\BibitemShut {NoStop}%
\bibitem [{\citenamefont {Tran}\ and\ \citenamefont {Blaha}(2009)}]{mbj}%
  \BibitemOpen
  \bibfield  {author} {\bibinfo {author} {\bibfnamefont {F.}~\bibnamefont
  {Tran}}\ and\ \bibinfo {author} {\bibfnamefont {P.}~\bibnamefont {Blaha}},\
  }\href@noop {} {\bibfield  {journal} {\bibinfo  {journal} {Phys. Rev. Lett.}\
  }\textbf {\bibinfo {volume} {102}},\ \bibinfo {pages} {226401} (\bibinfo
  {year} {2009})}\BibitemShut {NoStop}%
\bibitem [{\citenamefont {Koller}\ \emph {et~al.}(2011)\citenamefont {Koller},
  \citenamefont {Tran},\ and\ \citenamefont {Blaha}}]{koller}%
  \BibitemOpen
  \bibfield  {author} {\bibinfo {author} {\bibfnamefont {D.}~\bibnamefont
  {Koller}}, \bibinfo {author} {\bibfnamefont {F.}~\bibnamefont {Tran}}, \ and\
  \bibinfo {author} {\bibfnamefont {P.}~\bibnamefont {Blaha}},\ }\href@noop {}
  {\bibfield  {journal} {\bibinfo  {journal} {Phys. Rev. B}\ }\textbf {\bibinfo
  {volume} {83}},\ \bibinfo {pages} {195134} (\bibinfo {year}
  {2011})}\BibitemShut {NoStop}%
\bibitem [{\citenamefont {Singh}(2010{\natexlab{a}})}]{singh1}%
  \BibitemOpen
  \bibfield  {author} {\bibinfo {author} {\bibfnamefont {D.~J.}\ \bibnamefont
  {Singh}},\ }\href@noop {} {\bibfield  {journal} {\bibinfo  {journal} {Phys.
  Rev. B}\ }\textbf {\bibinfo {volume} {82}},\ \bibinfo {pages} {155145}
  (\bibinfo {year} {2010}{\natexlab{a}})}\BibitemShut {NoStop}%
\bibitem [{\citenamefont {Kim}\ \emph {et~al.}(2010)\citenamefont {Kim},
  \citenamefont {Marsman}, \citenamefont {Kresse}, \citenamefont {Tran},\ and\
  \citenamefont {Blaha}}]{kim-mbj}%
  \BibitemOpen
  \bibfield  {author} {\bibinfo {author} {\bibfnamefont {Y.~S.}\ \bibnamefont
  {Kim}}, \bibinfo {author} {\bibfnamefont {M.}~\bibnamefont {Marsman}},
  \bibinfo {author} {\bibfnamefont {G.}~\bibnamefont {Kresse}}, \bibinfo
  {author} {\bibfnamefont {F.}~\bibnamefont {Tran}}, \ and\ \bibinfo {author}
  {\bibfnamefont {P.}~\bibnamefont {Blaha}},\ }\href@noop {} {\bibfield
  {journal} {\bibinfo  {journal} {Phys. Rev. B}\ }\textbf {\bibinfo {volume}
  {82}},\ \bibinfo {pages} {205212} (\bibinfo {year} {2010})}\BibitemShut
  {NoStop}%
\bibitem [{\citenamefont {Singh}(2010{\natexlab{b}})}]{singh2}%
  \BibitemOpen
  \bibfield  {author} {\bibinfo {author} {\bibfnamefont {D.~J.}\ \bibnamefont
  {Singh}},\ }\href@noop {} {\bibfield  {journal} {\bibinfo  {journal} {Phys.
  Rev. B}\ }\textbf {\bibinfo {volume} {82}},\ \bibinfo {pages} {205102}
  (\bibinfo {year} {2010}{\natexlab{b}})}\BibitemShut {NoStop}%
\bibitem [{\citenamefont {Hicks}\ and\ \citenamefont
  {Dresselhaus}(1993)}]{hicks}%
  \BibitemOpen
  \bibfield  {author} {\bibinfo {author} {\bibfnamefont {L.~D.}\ \bibnamefont
  {Hicks}}\ and\ \bibinfo {author} {\bibfnamefont {M.~S.}\ \bibnamefont
  {Dresselhaus}},\ }\href@noop {} {\bibfield  {journal} {\bibinfo  {journal}
  {Phys. Rev. B}\ }\textbf {\bibinfo {volume} {47}},\ \bibinfo {pages} {12727}
  (\bibinfo {year} {1993})}\BibitemShut {NoStop}%
\bibitem [{\citenamefont {Chen}\ \emph {et~al.}(2013)\citenamefont {Chen},
  \citenamefont {Parker},\ and\ \citenamefont {Singh}}]{chen}%
  \BibitemOpen
  \bibfield  {author} {\bibinfo {author} {\bibfnamefont {X.}~\bibnamefont
  {Chen}}, \bibinfo {author} {\bibfnamefont {D.}~\bibnamefont {Parker}}, \ and\
  \bibinfo {author} {\bibfnamefont {D.~J.}\ \bibnamefont {Singh}},\ }\href@noop
  {} {\bibfield  {journal} {\bibinfo  {journal} {Sci. Rep.}\ }\textbf {\bibinfo
  {volume} {3}},\ \bibinfo {pages} {3168} (\bibinfo {year} {2013})}\BibitemShut
  {NoStop}%
\bibitem [{\citenamefont {Hornbostel}\ \emph {et~al.}(1997)\citenamefont
  {Hornbostel}, \citenamefont {Hyer}, \citenamefont {Thiel},\ and\
  \citenamefont {Johnson}}]{hornbostel}%
  \BibitemOpen
  \bibfield  {author} {\bibinfo {author} {\bibfnamefont {M.~D.}\ \bibnamefont
  {Hornbostel}}, \bibinfo {author} {\bibfnamefont {E.~J.}\ \bibnamefont
  {Hyer}}, \bibinfo {author} {\bibfnamefont {J.}~\bibnamefont {Thiel}}, \ and\
  \bibinfo {author} {\bibfnamefont {D.~C.}\ \bibnamefont {Johnson}},\
  }\href@noop {} {\bibfield  {journal} {\bibinfo  {journal} {J. Am. Chem.
  Soc.}\ }\textbf {\bibinfo {volume} {119}},\ \bibinfo {pages} {2665} (\bibinfo
  {year} {1997})}\BibitemShut {NoStop}%
\end{thebibliography}
%

\section*{Figure Captions}

\pagebreak

\end{document}